\documentclass[runningheads]{llncs}
\usepackage[english]{babel}
\usepackage[utf8]{inputenc}
\usepackage{microtype}
\usepackage{graphicx}
\usepackage{listings}
\usepackage{booktabs}
\let\accentvec\vec

\let\vec\accentvec

\newcommand{\x}{\textnormal{\textbf{x}}}

\usepackage{enumitem}

\usepackage{amsmath}

\usepackage{amsthm}
\usepackage{amssymb}
\usepackage{amsfonts}
\usepackage{verbatim}
\usepackage[textsize=tiny]{todonotes}
\usepackage{algorithm}
\usepackage[noend]{algpseudocode}

\usepackage{pgfplots}
\usetikzlibrary{patterns}
\usepgfplotslibrary{groupplots,colorbrewer}
\pgfplotsset{
  legend style = {font=\ttfamily}
}
\pgfplotsset{
cycle list/Set1-5,
cycle multiindex* list={
mark list*\nextlist
Set1-5\nextlist
},
}

\usepackage{url}

\definecolor{gruen}{rgb}{0.95,0.95,0.95}
\definecolor{dunkelgruen}{rgb}{0.35, 0.7, 0.67}
\definecolor{hellgruen}{rgb}{0.84, 0.7, 0.39}

\date{}

\title{The Role of Local Intrinsic Dimensionality in Benchmarking Nearest Neighbor Search}
\titlerunning{The Role of LID in Benchmarking Nearest Neighbor Search}

\author{
  Martin Aumüller\orcidID{0000-0002-7212-6476}
  \and 
  Matteo Ceccarello\orcidID{0000-0003-2783-0218}
}
\institute{IT University of Copenhagen\\ Copenhagen, Denmark\\ \email{\{maau,mcec\}@itu.dk}}
\begin{document}
\maketitle
\begin{abstract}
    This paper reconsiders common benchmarking approaches to nearest neighbor
    search. It is shown that the concept of local intrinsic dimensionality (LID)
    allows to choose query sets of a wide range of difficulty for real-world
    datasets. Moreover, the effect of different LID distributions on the running time
    performance of  implementations is empirically studied. 
    To this end, different visualization
    concepts are introduced that allow to get a more fine-grained overview of
    the inner workings of nearest neighbor search principles. 
    The paper closes with remarks about the diversity of datasets commonly used
    for nearest neighbor search benchmarking. 
    It is shown that
    such real-world datasets are not diverse: results on a single dataset
    predict results on all other datasets well.
\end{abstract}

\section{Introduction}

Nearest neighbor (NN) search is a key primitive in many computer science
applications, such as data mining, machine
learning and image processing.  For example, Spring and Shrivastava very
recently showed
in~\cite{Anshu17} how nearest neighbor search
methods can yield large speed-ups when training neural network
models. In this paper, we study the classical $k$-NN problem.
Given a dataset $S \subseteq \mathbb{R}^d$, the task is to build an index on $S$ to support the
following type of query: For a query point $\x\in\mathbb{R}^d$, return the $k$
closest points in $S$ under some distance measure $D$.

In many practical settings, a dataset consists of points represented as
high-dimensional vectors. For example, word representations generated by the
\texttt{glove} algorithm~\cite{pennington2014glove} associate with each word in a corpus a
$d$-dimensional real-valued vector. Common choices for $d$ are between 50 and 300 dimensions. 
Finding the true nearest neighbors in 
such a high-dimensional space is difficult, a phenomenon often referred to as the
``curse of dimensionality''\ \cite{Chavez01}. In practice, it means that finding
the true nearest neighbors, in general, cannot be solved much more efficiently
than by a linear scan through the dataset (requiring time $O(n)$ for $n$ data
points) or in space that is exponential in the dimensionality $d$, which is
impractical for large values of $d$.

While we cannot avoid these general hardness results~\cite{AlmanW15}, most datasets
that are used in applications are not \emph{truly} high-dimensional. This
means that the dataset can be embedded onto a lower-dimensional space without
too much distortion. Intuitively, the intrinsic dimensionality (ID) of the
dataset is the minimum number of dimensions that allows for such a
representation~\cite{Houle13}. There exist many explicit ways of finding good
embeddings for a given dataset.  For example, the Johnson-Lindenstrauss
transformation~\cite{JohnsonL86} allows us to embed $n$ data points in $\mathbb{R}^d$ into
$\Theta((\log n)/\varepsilon^2)$ dimensions such that all pairwise distances are
preserved up to a $(1+\varepsilon)$ factor with high probability. Another
classical embedding often employed in practice is given by principal component analysis (PCA), see~\cite{jolliffe2011principal}.

In this paper, we put our focus on ``local intrinsic dimensionality'' (LID),
a measure introduced by Houle in~\cite{Houle13}. We defer a
detailed discussion of this measure and its estimation to
Section~\ref{sec:measures}. Intuitively, the LID of a data point $\x$ at a distance threshold
$r > 0$ measures how difficult it is to distinguish between points
at distance $r$ and distance $(1+\varepsilon)r$ in a dataset. Most importantly for this
study, LID is a \emph{local} measure that can be associated with a single query. 
It was stated in~\cite{HouleSZ18} that the LID might serve as a 
characterization of the difficulty of $k$-NN queries. One purpose of this paper
is to shed light on this statement.

A focus of this paper is an empirical study of how the LID influences the performance of
NN algorithms. To be precise, we will benchmark four different
implementations~\cite{KriegelSZ17}
which employ different approaches to NN search. 
Three of them (\texttt{HNSW}~\cite{hnsw}, \texttt{FAISS-IVF}~\cite{faiss},
\texttt{Annoy}~\cite{annoy}) stood out as most performant in the empirical study
conducted by Aumüller et
al. in~\cite{AumullerBF17}. Another one (\texttt{ONNG}) was proposed very
recently~\cite{Iwasaki18} and shown to be competitive to these approaches. 
We base our experiments on~\cite{AumullerBF17} and describe their benchmarking
approach and the changes we made to their system in Section~\ref{sec:framework}. 
We analyze the LID distribution of real-world
datasets in Section~\ref{sec:algorithms:datasets}. We will see that there
is a substantial difference between the LID distributions
among 
datasets. We will next conduct two experiments: First, we fix a dataset and
choose as query set the set of points with smallest, medium, and largest
estimated LIDs. In addition, we choose a set of ``diverse'' query points w.r.t.
their LID estimates. 
As we will see, there is a clear tendency such that the larger the LID, the more difficult the query
for all implementations. Next, we will study how the different LID distributions
between datasets influence the running time performance. In a nutshell, it
cannot be concluded that LID by itself is a good indicator for the relative
performance of a fixed implementation over datasets. These statements will be
made precise in the evaluation that is discussed in Section~\ref{sec:evaluation}.

In the first part of our evaluation, we work in the ``classical evaluation
setting of nearest neighbor search''. This means that we relate a performance
measure (such as the achieved throughput measured in queries per second) to a quality
measure (such as the average fraction of true nearest neighbors found over all
queries). While this is the most commonly employed evaluation method, we reason 
that this way
of representing results in fact hides interesting details about the inner
workings of an implementation.  Using non-traditional visualization techniques
provide new insights into their query behavior on real-world datasets.  As one
example, we see that reporting average recall on the graph-based approaches
from~\cite{hnsw,Iwasaki18}  hides an important detail: For a given query, they
either find all true nearest neighbors or not a single one.  This behavior is
not shared by the two other approaches that we consider; both yield a continuous
transition from ``finding no nearest neighbors'' to ``finding all of them''.

As a final point, we want, ideally, to benchmark on a collection of
``interesting'' datasets that show the strengths and weaknesses of individual
approaches~\cite{smithmiles14}. We will
conclude that there is little diversity among the considered real-word datasets: While
the individual performance observations change from dataset to dataset, the relative
performance between implementations stays the same.

\paragraph{Our Contributions.}

The main contributions of this paper are

\begin{itemize}
    \item a detailed evaluation of the LID distribution of many real-world
        datasets used in benchmarking frameworks,
    \item an evaluation of the influence of the LID on the performance of NN
        search implementations,
    \item considerations about the result diversity, and
    \item an exploration of different visualization techniques that shed light
        on individual properties of certain implementation principles.
\end{itemize}

A preliminary workshop version of this paper appeared as~\cite{AumullerC19}. In this 
paper we expand the experimental study with the correlation between LID and recall;
we also consider different ways of generating synthetic datasets to investigate the
relationship between LID and performance.

\paragraph{Related Work on Benchmarking Frameworks for NN.}
We use the benchmarking system described in~\cite{AumullerBF17} as the starting
point for our study. Different approaches to benchmarking nearest neighbor
search are described in~\cite{mlpack2013,edel2014automatic,LiZSWZL16}. We refer
to~\cite{AumullerBF17} for a detailed comparison between the frameworks. 

\section{Local Intrinsic Dimensionality}\label{sec:measures}

We consider a distance-space $(\mathbb{R}^d, D)$ with a distance function
$D\colon\mathbb{R}^d \times \mathbb{R}^d \rightarrow \mathbb{R}$.
As described in~\cite{Amsaleg15}, we consider the distribution of distances
within this space with respect to a reference point $\x$. Such a distribution
is induced by sampling $n$ points from the space $\mathbb{R}^d$ under a certain
probability distribution. We let $F\colon\mathbb{R} \rightarrow [0,1]$ be the
cumulative distribution function of distances to the reference point $\x$.

\begin{definition}[\cite{Houle13}]
The local continuous intrinsic dimension of $F$ at distance $r$ is given by
\begin{equation*}
    \text{ID}_F(r) = \lim_{\varepsilon\rightarrow
    0}\frac{\ln(F((1+\varepsilon)r)/F(r))}{\ln((1+\varepsilon)r/r)},
\end{equation*}
whenever this limit exists.
\end{definition}
The measure relates the increase in distance to the increase in probability mass (the
fraction of points that are within the ball of radius $r$ and $(1+\varepsilon)r$
around the query point). Intuitively, the larger the LID, the more
difficult it is to distinguish true nearest neighbors at distance $r$ from the rest of the
dataset. As described in~\cite{HouleSZ18}, in the context of
$k$-NN search we set $r$ as the distance of the $k$-th nearest neighbor to the
reference point $\x$. 

\paragraph{Estimating LID} We use the
Maximum-Likelihood estimator (MLE) described in~\cite{Levina05,Amsaleg15} to
estimate the LID of $\x$ at distance $r$. Let $r_1
\leq \ldots \leq r_k$ be the sequence of distances of the $k$-NN of 
$\x$. The
MLE $\hat{\text{ID}}_\x$ is then 
\begin{equation}
    \hat{\text{ID}}_\x = -\left(\frac{1}{k}\sum_{i = 1}^{k} \ln \frac{r_i}{r_k}\right)^{-1}.
    \label{eq:lid:estimate}
\end{equation}
Amsaleg et al. showed in \cite{Amsaleg15} that MLE estimates the LID well. We remark that in very recent work, Amsaleg et al. proposed in~\cite{amsaleg2019intrinsic} a new MLE-based estimator that works with smaller $k$ values than~\eqref{eq:lid:estimate}.

\section{Overview over the Benchmarking Framework}\label{sec:framework}
We use the \texttt{ann-benchmarks} system described in~\cite{AumullerBF17} to
conduct our experimental study. Ann-benchmarks is a framework for benchmarking
NN search algorithms. It covers dataset creation, performing the actual 
experiment, and storing the results of these experiments in a transparent and
easy-to-share way. Moreover, results can be explored 
through various plotting functionalities, e.g., by creating a website
containing interactive plots for all experimental runs.

Ann-benchmarks interfaces with a NN search implementation by calling its 
preprocess (index building) and search (query) methods with certain parameter
choices. Implementations are tested on a large set of parameters
usually provided by the original authors of an implementation.
The answers to queries are recorded as the indices
of the points returned.  
Ann-benchmarks stores these parameters together with further statistics such as individual query times, index size, and auxiliary information
provided by the implementation. 
See~\cite{AumullerBF17} for more details. 

Compared to the system described in~\cite{AumullerBF17}, we
added tools to estimate the LID based on Equation~\eqref{eq:lid:estimate}, pick ``challenging
query sets'' according to the LID of individual points, and added new datasets and
implementations. Moreover, we implemented a mechanism that
allows an implementation to provide further query statistics after answering a
query. To showcase this feature, all implementations in this study 
report the number of distance computations performed to answer a
query.\footnote{We thank the authors of the implementations
for their help and responsiveness in adding this feature to their
library.} 

    
\section{Algorithms and Datasets}\label{sec:algorithms:datasets}
\begin{table*}[t]
  \small
  \centering
  \begin{tabular}{l r @{\hskip 0.5em} r@{\hskip 1.5em} r@{\hskip 0.5em} r @{\hskip 1em} l}
    \toprule
     &  &  & \multicolumn{2}{c}{\textbf{LID}} & \\
     \cmidrule{4-5}
    \textbf{Dataset} & \textbf{Data Points} & \textbf{Dimensions} & \textbf{avg} & \textbf{median} & \textbf{Metric} \\
    \midrule
    \textsf{SIFT}~\cite{JegouDS11} & 1\,000\,000 & 128 & 21.9 & 19.2 & Euclidean \\
    \textsf{MNIST} & \phantom{0\,0}65\,000 & 784 & 14.0  & 13.2 & Euclidean \\
    \textsf{Fashion-MNIST}~\cite{fashion-mnist} & \phantom{0\,0}65\,000 & 784 & 15.6  & 13.9 & Euclidean \\
    \textsf{GLOVE}~\cite{pennington2014glove} & 1\,183\,514 & 100 & 18.0 & 17.8 & Angular/Cosine \\
    \textsf{GLOVE-2M}~\cite{pennington2014glove}& 2\,196\,018 & 300 & 26.1 & 23.4 & Angular/Cosine \\
    \textsf{GNEWS}~\cite{word2vec} & 3\,000\,000 & 300 & 21.1 & 20.1 & Angular/Cosine \\
    \bottomrule
  \end{tabular}
\caption{Datasets under consideration with their average local intrinsic
    dimensionality (LID) computed by MLE~\cite{Amsaleg15} from the 100-NN of all
    the data points. 
}
\label{tab:datasets}
\end{table*}

\subsection{Algorithms}

Nearest neighbor search algorithms for high dimensions are usually graph-,
tree-, or hashing-based. We refer the reader to~\cite{AumullerBF17} for an
overview over these principles and available implementations. 
In this study, we concentrate on the three implementations
considered most performant in~\cite{AumullerBF17}, namely
\texttt{HNSW}~\cite{hnsw}, \texttt{Annoy}~\cite{annoy} and
\texttt{FAISS-IVF}~\cite{faiss} (\texttt{IVF} from now on). We consider the very recent graph-based
approach \texttt{ONNG}~\cite{Iwasaki18} in this study as well.

\texttt{HNSW} and \texttt{ONNG} are graph-based approaches. This means that they
build a $k$-NN graph during the preprocessing step. In this graph, each vertex
is a data point and a directed edge $(u, v)$ means that the
data point associated with $v$ is ``close'' to the data point associated with
$u$ in the dataset. At query time, the graph is traversed to generate candidate
points. Algorithms differ in details of the graph construction, how they build a navigation structure on top
of the graph, and how the graph is traversed.

\texttt{Annoy} is an implementation of a random projection forest, which is a
collection of random projection trees. 
Each node in a tree is associated with a set of data points. It splits these
points into two subsets according to a
chosen hyperplane.
If the dataset in a node is small enough, it is stored directly and the node is
a leaf.
\texttt{Annoy} employs a data-dependent splitting mechanism 
in which a splitting hyperplane is chosen as the one splitting two ``average points''
by repeatedly sampling dataset points. In the query phase, trees are traversed using a
priority queue until a predefined number of points is found.  

\texttt{IVF} builds an inverted file based on clustering the dataset around
a predefined number of centroids. It splits the dataset based on these centroids by associating each point with its closest centroid. During query it finds the closest centroids
and checks points in the dataset associated with those.

We remark we used both \texttt{IVF} and \texttt{HNSW} implementations from 
\texttt{FAISS}\footnote{\url{https://github.com/facebookresearch/faiss}}.

\subsection{Datasets}
Table~\ref{tab:datasets} presents an overview over the datasets that we consider
in this study.
We restrict our attention to datasets that are usually used in
connection with Euclidean distance and Angular/Cosine distance.
For each dataset, we compute the LID distribution with respect to the
100-NN as discussed in Section~\ref{sec:measures}. \textsf{SIFT},
\textsf{MNIST}, and \textsf{GLOVE} are among the most-widely used
datasets for benchmarking nearest neighbor search algorithms.
\textsf{Fashion-MNIST} is considered as a replacement for \textsf{MNIST},
which is usually considered too easy for machine learning
tasks~\cite{fashion-mnist}.

\begin{figure}[t]
    \begin{minipage}[c]{.55\textwidth}
  \includegraphics[width=\columnwidth]{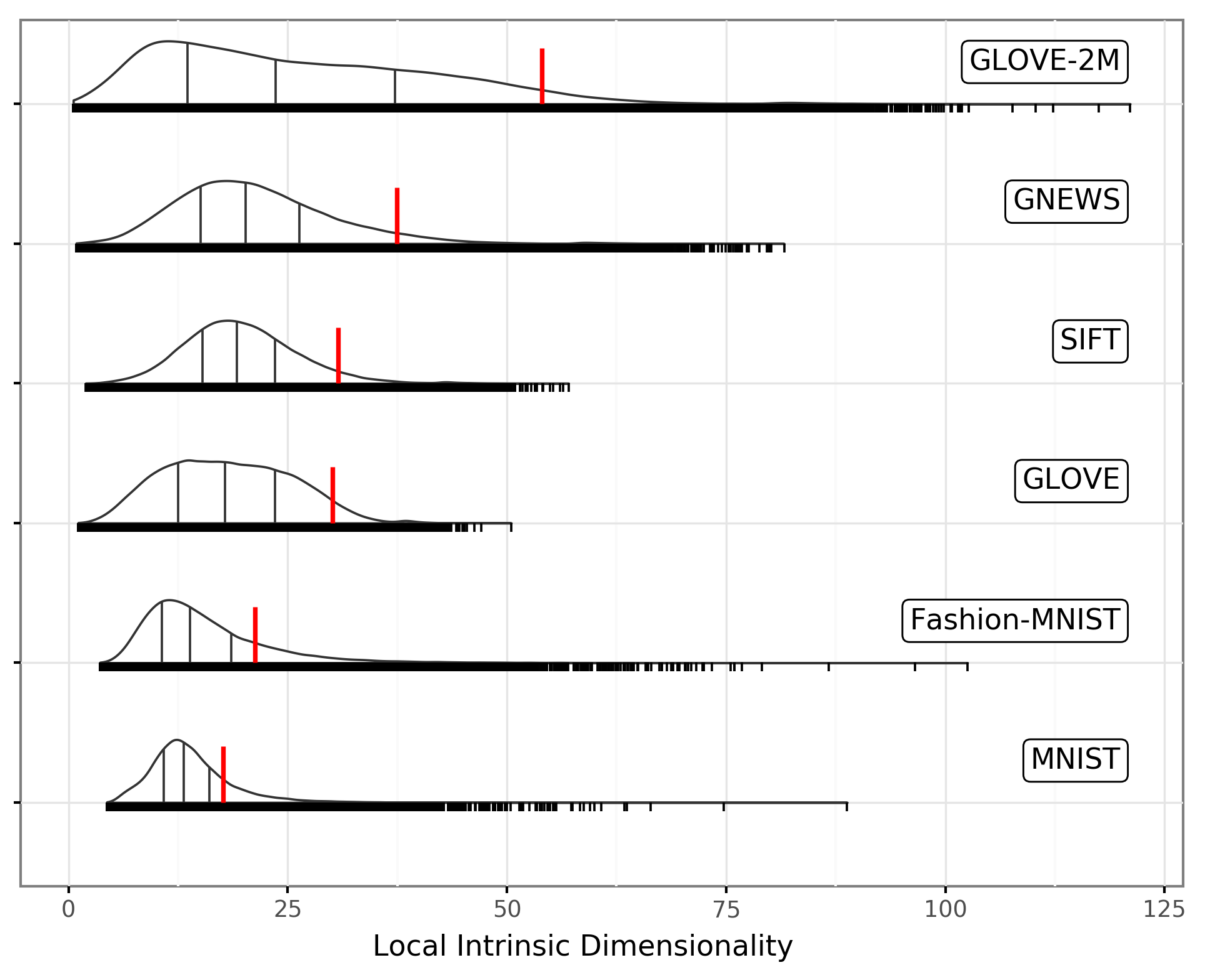}
  \end{minipage}\hfill
  \begin{minipage}[c]{.4\textwidth}
  \caption{LID distribution for each dataset.
  Ticks below the distribution curves represent single queries.
  Lines within each distribution curve correspond to the 25, 50 and 75
percentile. The red line marks the 10\,000 largest estimated LID, which we use as a threshold value to define \emph{hard} query sets.}
    \label{fig:datasets:lid}
\end{minipage}
\end{figure}

Figure~\ref{fig:datasets:lid} provides a visual representation of the 
estimated LID distribution of each dataset, for $k=100$.
While the datasets differ widely in their original
dimensionality, the median LID ranges from around 13 for \textsf{MNIST} to about 23 for 
\textsf{GLOVE-2M}. The distribution of LID values is asymmetric and shows a 
long tail behavior. \textsf{MNIST}, \textsf{Fashion-MNIST}, \textsf{SIFT}, and
\textsf{GNEWS} are much more concentrated around the median compared to the two
\texttt{glove}-based datasets.

\section{Evaluation}\label{sec:evaluation}
This section reports on the results of our experiments. Due to space
constraints, we only present some selected results. More results and plots
can be explored via interactive
plots at \url{http://ann-benchmarks.com/sisap19/}, which also contains a link to
the source code repository. For a fixed implementation, the plots presented here
consider the Pareto frontier over all parameter choices~\cite{AumullerBF17}. Tested parameter choices
and the associated plots are available on the website. 

\paragraph{Experimental Setup}
Experiments were run on 2x 14-core Intel Xeon E5-2690v4 (2.60GHz) with 512GB RAM using 
Ubuntu 16.10 (kernel 4.4.0). Index building was multi-threaded, queries 
where answered in a single thread.

\paragraph{Quality and Performance Metrics} As quality metric we measure the
individual recall of each query, i.e., the fraction of points reported by the
implementation that are among the true $k$-NN. As performance metric, we record 
individual query times and the total number of distance computations needed to
answer all queries. We usually report on the throughput (the average number of queries that can
be answered in one second, in the plots denoted as QPS for \emph{queries per second}), but we will also inspect individual query times. 

\paragraph{Objectives of the Experiments} 
Our experiments are tailored to answer the following questions:

\begin{enumerate}
    \item[(Q1)] How does the LID of a query set influence the running time
      performance? (Section~\ref{sec:lid-influence})
    \item[(Q2)] How diverse are measurements obtained on datasets? Do relative
        differences between the performance of different implementations stay
        the same over multiple datasets? (Section~\ref{sec:results-diversity})
    \item[(Q3)] How well does the number of distance computations reflect the 
        relative running time performance of the tested implementations? (Section~\ref{sec:results-diversity})
    \item[(Q4)] How concentrated are quality and performance measures around
        their mean for the tested implementations?
        (Section~\ref{sec:uncertainty}) 
\end{enumerate}

\begin{figure*}[t]
%
%
\input{plot-running-time-lid}
\end{figure*}
\begin{figure*}[t]

\input{plot-all-datasets}
\end{figure*}

\paragraph{Choosing Query Sets} For each dataset, we 
select four different query sets: The query set that
contains the 10\,000 points with the lowest estimated LID (which we denote \emph{easy}), 10\,000 points around the data point with median estimated LID
(denoted \emph{medium}), 10\,000 points with the largest estimated LID
(dubbed \emph{hard}), and 5\,000 points chosen uniformly according to (integer) LID values (denoted \emph{diverse}).
For the latter, we split all data points up into buckets, where bucket $i$ represents all data points that have 
an estimated LID of $i$ (rounded down). For each query, we pick a non-empty bucket uniformly at random, and inside
the bucket we pick a random point (with repetition).
Figure~\ref{fig:datasets:lid} marks with a red line the LID used as a threshold to build the \emph{hard} queryset.

%
%

\subsection{Influence of LID on Performance}
\label{sec:lid-influence}

Figure~\ref{plot:lid:running:time} shows results for the influence of using
only points with low, middle, and large estimated LID as query points,
in \textsf{SIFT} and \textsf{GLOVE-2M}. 
We observe a clear influence of the LID of the query set on the
performance: the larger the LID, the more down and to the left the graphs move.
This means that, for higher LID, it is more expensive, in terms of time, to
answer queries with good recall.
For all datasets except \textsf{GLOVE-2M}, all implementations were still able to
achieve close to perfect recall with the parameters set. This means that all but
one of the tested datasets do not contain too many ``noisy queries''. Already the queries
around the median prove challenging for most implementations. For the most
difficult queries (according to LID), only \texttt{IVF} and \texttt{ONNG}
achieve close to perfect recall on \textsf{GLOVE-2M}. 

Figure~\ref{plot:datasets:running:time} reports on the results of \texttt{ONNG}
and \texttt{Annoy} on selected datasets. Comparing results to the LID
measurements depicted in Figure~\ref{fig:datasets:lid}, the estimated median LID
gives a good estimate on the relative performance of the algorithms on the data
sets. As an exception, \textsf{SIFT (M)} is much easier than predicted by its
LID distribution. In particular for \texttt{Annoy}, the hard \textsf{SIFT}
instance is as challenging as the medium \textsf{GLOVE} version.   The easy
version of \textsf{GLOVE-2M} turns out to be efficiently solvable by both
implementations (taking about the same time as it takes to answer the hard instance of
\textsf{Fashion-MNIST}, which has a much higher LID). From this, we cannot
conclude that LID as a single indicator explains performance differences
of an implementation across different datasets. However, more careful experimentation
is need before drawing a final conclusion. In our setting, the LID estimation is 
conducted for $k=100$, while queries are only searching for the 
10 nearest neighbors. Moreover, the estimation using MLE might not be accurate enough
on these datasets, since it is very dependent on the parameter $k$ being used.
We leave the investigation of these two directions as future work.

In general, the diverse query set is more difficult than the medium query set. In particular,
at high recall it generally becomes nearly as difficult as the difficult dataset. The reason for this behavior
is that none of the implementations can adapt to the difficulty of a query. They only achieve high average recall
when they can solve sufficiently many queries with high LID. The parameter settings that allow for such guarantees 
slow down answering the easy queries by a lot. We believe that the ``diverse'' query sets thus allow for 
challenging benchmarking datasets for adaptive query algorithms.

As a side note, we remark that \textsf{Fashion-MNIST} is as difficult to solve
as \textsf{MNIST} for all implementations, and is by far the easiest dataset
for all implementations. Thus, while there is a big difference in the difficulty
of solving the classification task~\cite{fashion-mnist}, there is no measurable difference between
these two datasets in the context of NN search.

\begin{figure}[t]
  \includegraphics[width=.49\columnwidth]{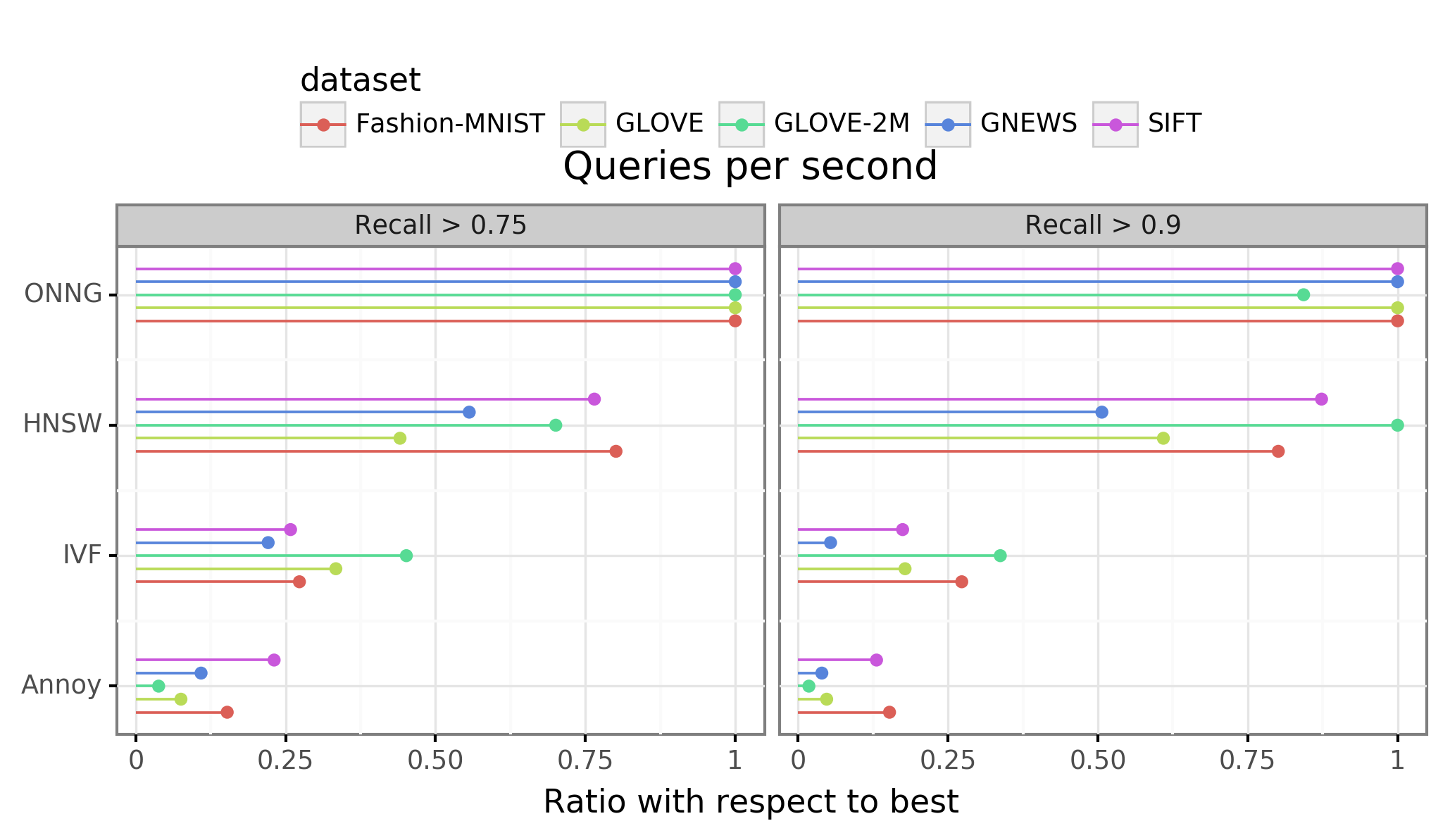}
  \hfill
  \includegraphics[width=.49\columnwidth]{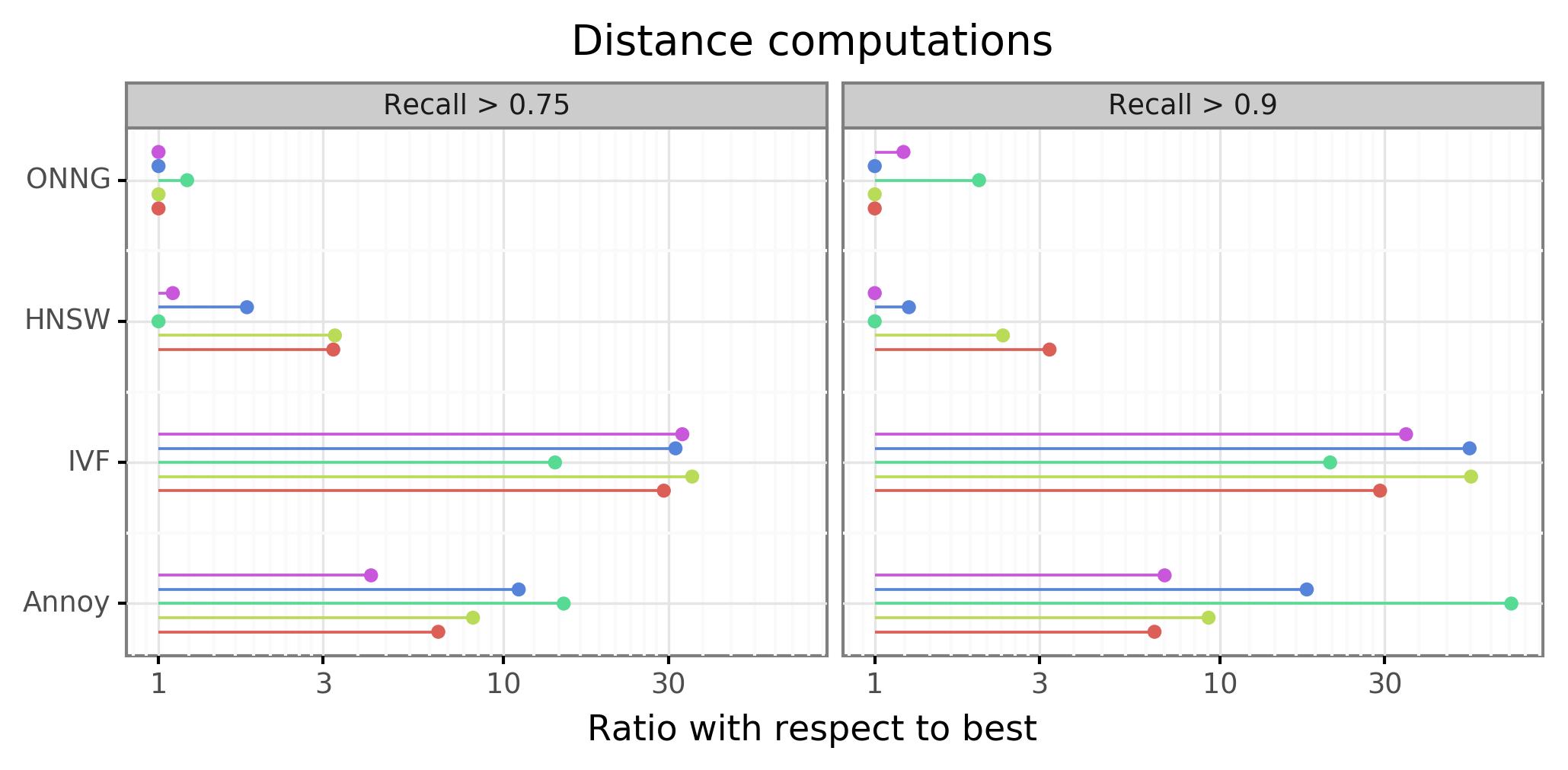}
  \caption{Ranking of algorithms on five different datasets, according to recall
  $\ge 0.75$ and $\ge 0.9$, and according to two different performance measures:
number of queries per second (left) and number of distance computations (right).
Both plots report the ratio with the best performing algorithm on each dataset:
for the queries per second metric a larger ratio is better, for the number of
distance computations metric, a smaller ratio is better.}
  \label{fig:ranking}
\end{figure}

\subsection{Diversity of Results}
\label{sec:results-diversity}

Figure~\ref{fig:ranking} gives an overview over how algorithms compare to each
other among all ``medium difficulty'' datasets.
We consider two metrics, namely the number of queries per second (left plot),
and the number of distance computations (right plot).
For two different average recall thresholds (0.75 and 0.9) we then select, for
each algorithm, the
best performing parameter configuration that attains at least that recall.
For each dataset, the plots report the ratio with the best performing
algorithm on that dataset, therefore the best performer is reported with ratio
1.
Considering different dataset, we see that there is little
variation in the ranking of the algorithms. Only the two graph-based approaches trade ranks, all
other rankings are stable. Interestingly, \texttt{Annoy} makes much fewer
distance computations but is consistently outperformed by
\texttt{IVF}.\footnote{We note that \texttt{IVF} counts the initial comparisons
to find the closest centroids as distance computations, whereas \texttt{Annoy}
did not count the inner product computations during tree traversal.} 

Comparing the number of distance computations to running time performance, we
see that an increase in the number of distance computations is not reflected in
a proportional decrease in the number of queries per second.
This means that the candidate set generation is in general more expensive for
graph-based approaches, but the resulting candidate set is of much higher
quality and fewer distance computations have to be carried out.
Generally, both graph-based algorithms are within a factor 2 from
each other, whereas the other two need much larger candidate lists to achieve a
certain recall. The relative difference usually ranges from 5x to 30x more distance
computations for the non-graph based approaches, in particular at high recall. 
This translates well into the performance differences we see in this setting:
consider for instance Figure~\ref{plot:lid:running:time}, where the lines corresponding to
\texttt{HNSW} and \texttt{ONNG} upper bound the lines relative to
the other two algorithms.

\subsection{Reporting the Distribution of Performance}
\label{sec:uncertainty}

\begin{figure*}
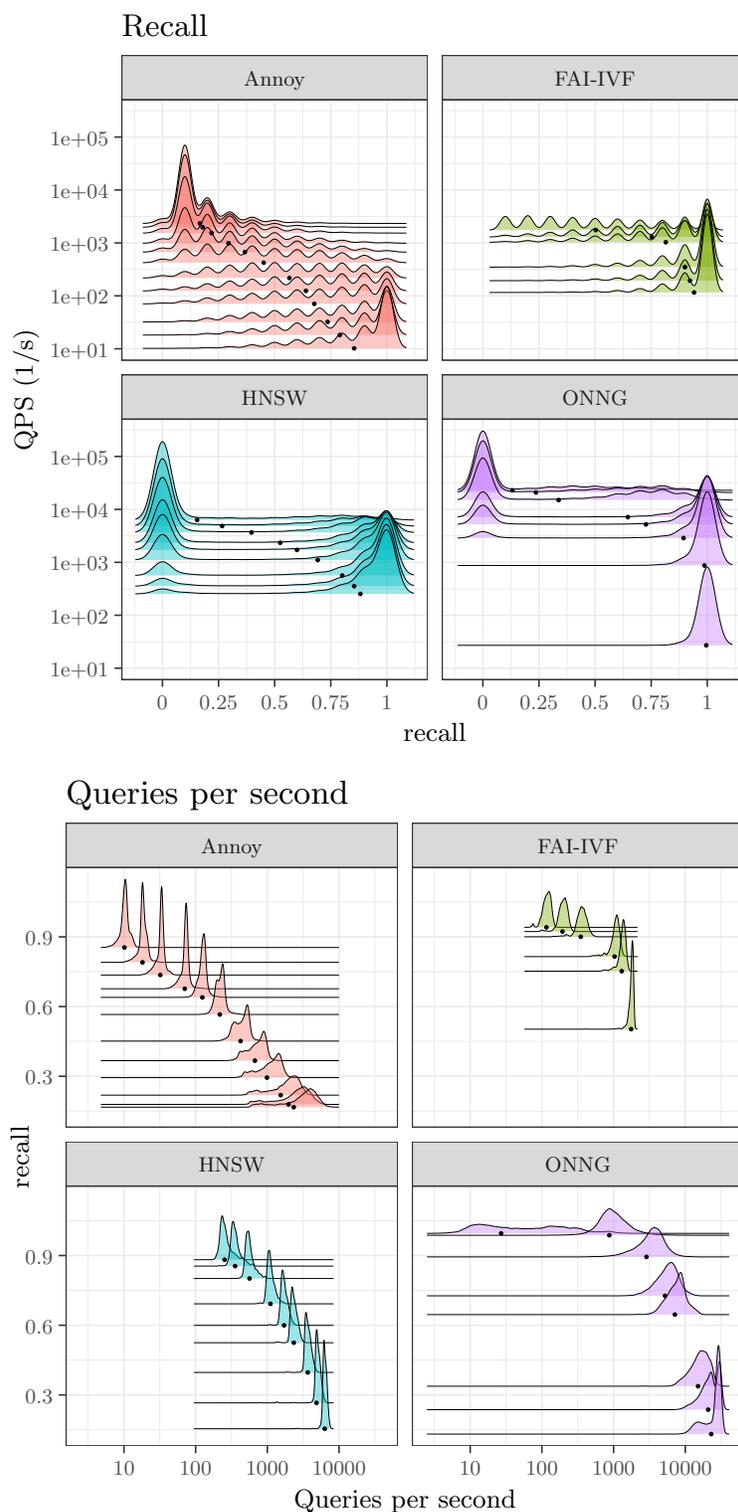

  \centering
    \input{imgs/recall_distribution.tex}
    \input{imgs/query_time_distribution.tex}

\vspace*{-2em}
    \caption{Distribution of performance for queries on the \textsf{GLOVE-2M}
    (medium difficulty) dataset. Looking just at the average performance can hide interesting behaviour.
  }
  \label{fig:uncertainty-plots}
\end{figure*}

\begin{figure*}
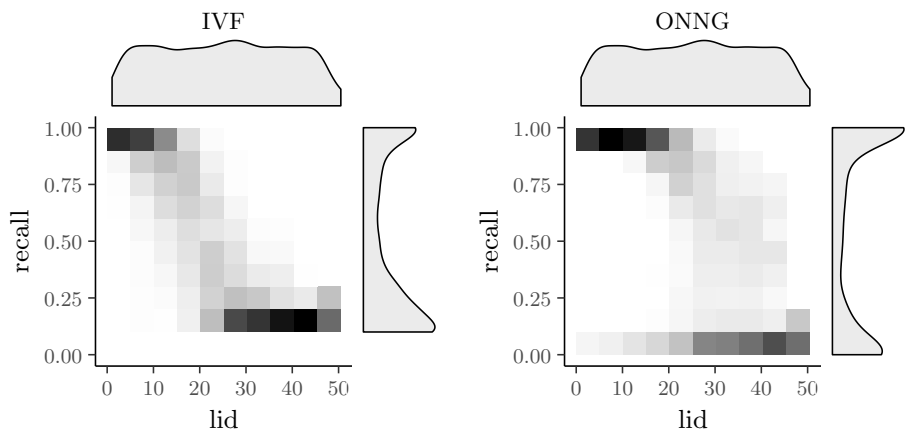

    \centering
    \begin{minipage}{.48\textwidth}
      \input{imgs/ivf-glove-lid-recall.tex}
    \end{minipage}
    \hfill
    \begin{minipage}{.48\textwidth}
      \input{imgs/onng-glove-lid-recall.tex}
    \end{minipage}
    \caption{Distribution of Recall vs. LID plot on the \textsf{GLOVE} datset. Intensity reflects number of queries that
    achieve a combination of recall vs. LID.}
    \label{fig:lid:recall}
\end{figure*}

In the previous sections, we made extensive use of recall/queries per second
plots, where each configuration of each algorithm results in a single point,
namely the average recall and the inverse of the average query time. As we
shall see in this section, concentrating on averages can hide interesting
information in the context of $k$-NN queries.
In fact, not all queries are equally difficult to answer.
Consider the plots in Figure~\ref{fig:uncertainty-plots},
which report performance of the four algorithms\footnote{In order not to
clutter the plots, we fixed parameters as follows: 
\texttt{IVF} | number of lists 8192;
\texttt{Annoy} | number of trees 100;
\texttt{HNSW} | efConstruction 500, M 8;
\texttt{ONNG} | edge 100, outdegree 10, indegree 120. 
} 
on the \textsf{GLOVE-2M} dataset, medium difficulty.
The top four plots report the recall versus the number of queries per second, and
black dots correspond to the averages. Additionally, for each
configuration, we report the distribution of the recall scores: the baseline of
each recall curve is positioned at the corresponding queries per second performance.
Similarly, the bottom plots report on the inverse of the individual query times 
(the average of these is the QPS in the left plot) against the average recall. 
In both plots, the best performance is achieved towards the top-right corner.
 
Plotting the distributions, instead of just reporting the averages, uncovers
some interesting behaviour that might otherwise go unnoticed, in particular with
respect to the recall.
The average recall gradually shifts towards the right as the effect of more and
more queries achieving good recalls.
Perhaps surprisingly, for graph-based algorithms this shift is very sudden: most
queries go from having recall 0 to having recall 1, taking no intermediate
values.
Taking the average recall as a performance metric is convenient in that it is a
single number to compare algorithms with. However, the same average recall can
be attained with very different distributions: looking at such distributions can
provide more insight.

For the bottom plots, we observe that individual query times 
of all the algorithms are well concentrated around their mean. 

Figure~\ref{fig:lid:recall} gives another distributional view on the achieved result quality.
The plots shows two runs of \texttt{IVF} and \texttt{ONNG} with fixed parameters on the 
\textsf{GLOVE} dataset with diverse queries. On the top we see the distribution of estimated 
LID values for the diverse query set, on the right we see the distribution of recall values achieved by the implementation. 
Each of the 
queries corresponds to a single data point in the recall/LID plot and data points are summarized 
through hexagons, where the color intensity of a hexagon indicates the number of data points
falling into this region. The plots show that the higher the LID of a query, there is a clear tendency for the query 
to achieve lower recall.   

For space reasons, we do not report other parameter configurations and datasets,
which nonetheless show similar behaviours. All of them can be accessed at the website.

\section{Summary}

In this paper we studied the influence of LID to the performance of nearest
neighbor search algorithms. We showed that LID allows to choose query sets of a
wide range of difficulty from a given dataset. We also showed how different LID
distributions influence the running time performance of the algorithms. In this respect, we
could not conclude that the LID alone can predict running time differences well.
In particular, \textsf{SIFT} is usually easier for the algorithms, while
\textsf{GLOVE}'s LID distribution would predict it to be the easier dataset of
the two.

With regard to challenging query workloads, we described a way to choose diverse 
query sets. They have the property that for most implementations it is easy to perform
well for most of the query points, but they contain many more easy and difficult queries
than query workloads chosen randomly from the dataset. We believe this is a very interesting
benchmarking workload for approaches that try to adapt to the difficulty of an individual query.

We introduced novel visualization techniques to show the uncertainty within
the answer to a set of queries, which made it possible to show a clear
difference between the graph-based algorithms and the other
approaches. 

We hope that this study initiates the search for more diverse datasets, or for
theoretical reasoning why certain algorithmic principles are generally better
suited for nearest neighbor search. 
On a more practical side, Casanova et al. showed in~\cite{CasanovaEHKNSZ17} 
how dimensionality testing can be used to speed up reverse $k$-NN queries. We would be
interested in seeing whether the LID can be used at other places in the design of NN algorithms to
guide the search process or the parameter selection. While we know from~\cite{Amsaleg15} that the LID estimation
of MLE with $k = 100$ works well on their datasets, it would be interesting to see whether the other estimators 
mentioned there are also able to characterize the relative performance of queries.

\paragraph*{Acknowledgements} 
The authors would like to thank the anonymous reviewers for their useful suggestions, which helped to improve the presentation of the paper.  The research leading to these results has received funding from the European Research Council under the European Union's 7th Framework Programme (FP7/2007-2013) / ERC grant agreement no. 614331.

\bibliographystyle{splncs04}
\bibliography{lit}

\end{document}